\documentclass{chjaa}

\usepackage{graphicx,times}
\input{epsf.sty}
\input{psfig.sty}
\newcommand{\gsim}{\lower.5ex\hbox{$\; \buildrel > \over \sim \;$}}
\newcommand{\lsim}{\lower.5ex\hbox{$\; \buildrel < \over \sim \;$}}

\newcommand{\hii}{\hbox{H\,{\sc ii}}}
\newcommand{\heii}{\hbox{He\,{\sc ii}}}

\begin{document}

\title{Clustering Property of Wolf-Rayet Galaxies in the SDSS\,$^*$
\footnotetext{\small $*$ Supported by the National Natural Science
Foundation of China.} }

\volnopage{Vol.\ 8 (2008), No.\ 2,~ 211--218}
   \setcounter{page}{211}

\author{Wei Zhang
  \and
       Xu Kong
  \and
       Fu-Zhen Cheng
       }

\institute{Center for Astrophysics, University of Science and
Technology of China, Hefei 230026; {\it xtwfn@mail.ustc.edu.cn}\\
\vs\no
   {\small Received 2007 April 27; accepted 2007 May 24}
}

\abstract{ We have analysed, for the first time, the clustering properties  
of  Wolf-Rayet (W-R) galaxies, using a large sample of 846 W-R galaxies 
selected from the  Data Release 4 (DR4) of  the Sloan  Digital  Sky  Survey  
(SDSS).   We  compute  the cross-correlation function between  W-R  galaxies  
and  a reference sample  of galaxies drawn from the DR4.  We compare the 
function to the results for control samples of non-W-R star-forming galaxies
that are matched closely in redshift, luminosity, concentration, 4000-\AA\ 
break strength  and specific star formation rate (SSFR). On scales larger than 
a few Mpc, W-R galaxies  have almost the same clustering amplitude as the control 
samples, indicating that W-R galaxies and non-W-R control galaxies populate dark 
matter haloes of similar masses. On scales between 0.1--1$h^{-1}$~Mpc, W-R galaxies
are less clustered  than the control samples, and the size  of the difference 
depends on the SSFR. Based on both observational and theoretical considerations, 
we speculate  that this negative bias can be interpreted by W-R galaxies residing 
preferentially at the centers of their dark matter haloes. We examine the distribution  
of W-R galaxies more closely using the SDSS galaxy group catalogue of Yang et  al., 
and find that $\sim$82\% of  our W-R  galaxies are  the  central galaxies of groups, 
compared to $\sim$74\% for the corresponding control galaxies.  We  find that W-R 
galaxies are hosted, on average, by dark matter  haloes  of masses  of 
$10^{12.3}M_\odot$, compared to $10^{12.1}M_\odot$ for centrally-located W-R galaxies 
and $10^{12.7}M_\odot$ for satellite ones.  We would like to point out that this 
finding, which provides a direct observational support to our conjecture, is really 
very crude due to the small number  of W-R galaxies  and the incompleteness of 
the group catalogue, and needs more  work  in future with larger samples.
\keywords{galaxies: distances and redshifts  ---
          galaxies:  starburst  ---
          stars:     Wolf-Rayet}
}

\authorrunning{W. Zhang, X. Kong \&  F. Z. Cheng}
\titlerunning{The Clustering of Wolf-Rayet Galaxies}

\maketitle

\section{Introduction}

Wolf-Rayet (W-R) galaxies are a subset of emission line galaxies
whose integrated spectra show direct signatures
of
W-R stars, most commonly, a {\it broad} emission feature at 4686\,\AA\ \heii\
originating in the stellar winds (\cite{Conti91}). These galaxies
are thought to be undergoing present or very recent star formation
that produces massive  stars  evolving  to the W-R stage. This
indicates
typical ages of \lsim 10\,Myr  and
initial masses of \gsim 20\,$M_\odot$ (\cite{Maeder94}). They are therefore
ideal objects
for
studying the early phases of starbursts
and
the
burst properties, and
for
constraining
the parameters of the stellar initial mass function (IMF) (\cite{Schaerer99};
\cite{Guseva00}). Since the first detection of W-R features in He 2--10
(\cite{Allen76}),
$\sim$140 W-R galaxies
have been reported by the end of the last century,
some
found by
systematic searches, but most, serendipitously (\cite{Schaerer99}).
With the large redshift surveys
assembled in recent years, in particular the Sloan Digital Sky
Survey (SDSS; \cite{York00}), the number of W-R galaxies has grown
rapidly, and it has become possible to study these galaxies using
large and homogeneous  samples (\cite{Kniazev04}; \cite{Zhang07}).

We have presented an SDSS-based sample of 174 W-R  galaxies  in
Zhang et al. (2007), where the relationship between galaxy
metallicity and  IMF  slope  was  studied. Here we present a new
sample consisting of 846 W-R galaxies from the SDSS Data Release 4
(DR4, \cite{Adelman06}). This is, to date, the largest and most
homogeneous sample of W-R galaxies, which reveals that W-R
galaxies are not a random subsample of star-forming galaxies.
Rather, they are less luminous, younger in mean stellar age, bluer
in color, poorer in metal and  more concentrated in structure.  In
particular, W-R galaxies exhibit the highest specific star
formation rates (SSFRs). In this paper, we use  the new  catalogue
to extend our study of W-R galaxies to an especial analysis of
their clustering.

The clustering of galaxies as a function of  their  properties
provides strong tests for theoretical models of structure and galaxy
formation (e.g. \cite{Peebles80}).
The
clustering is usually quantified using the two-point correlation function (2PCF). In the standard model
of structure formation, the amplitude of the 2PCF on scales larger than a few Mpc provides a direct measure of the mass of the dark matter haloes that host the galaxies (\cite{Kaiser86}), while the slope of the 2PCF on small scales depends on the detailed
location
of the galaxies
inside/around those dark  matter haloes (\cite{Zehavi04}). In this paper, we present the two-point cross-correlation function between W-R galaxies and a reference sample of galaxies drawn from the DR4. This  is  the first
determination of the clustering of this special class of galaxies
and,  as we will see, the  data  are  now  large enough to allow for
a clustering measurement with acceptable accuracy. We wish to
isolate the effect of the W-R features, and so we compare the
measurement to that of control samples of non-W-R star-forming
galaxies that are closely matched in redshift, luminosity,
concentration, star formation rate (SFR) and mean stellar age, as
measured by the 4000-\AA\ break strength (D$_{4000}$). As pointed
out by Li et al. (2006b), this close matching is important because
previous work has established  that the clustering of galaxies
depends strongly on these properties (e.g. \cite{Zehavi05};
\cite{Li06a}). With the help of closely-matched control samples, we
will be able to understand whether there is a real physical
connection between the location of a galaxy and the W-R features
found in it.

Throughout this paper, we assume a cosmological model with the
density parameter $\Omega_0=0.3$  and cosmological constant
$\Lambda_0=0.7$. A Hubble constant $h=1$, in units of
100~km~s$^{-1}$~Mpc$^{-1}$, is assumed when computing
the absolute magnitudes.

\section{Data and Methodology}

\begin{figure}
\centering
\includegraphics[width=0.6\columnwidth]{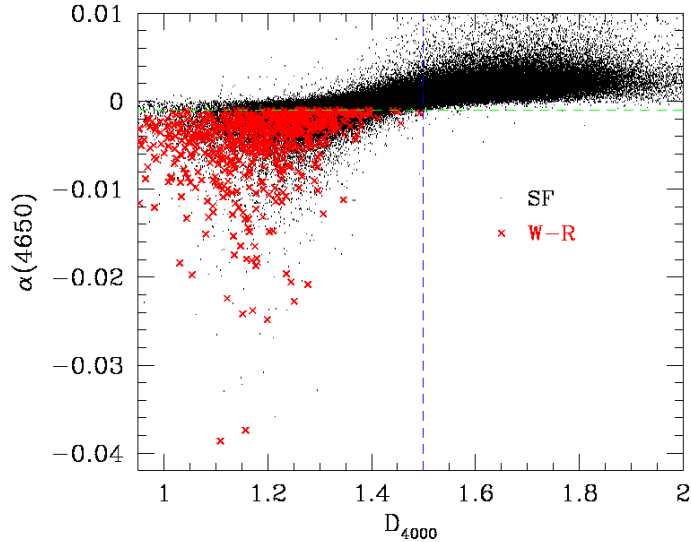}
\caption{\baselineskip 3.6mm Distribution of star-forming galaxies
(dots) and W-R galaxies (times signs) in the ($\alpha$(4650)
--D$_{4000}$) diagram. Horizontal  dashed line marks
$\alpha(4650)=-0.001$, vertical  dashed  line, D$_{4000}=1.5$.
Note that for all the  W-R galaxies, we have D$_{4000}$ smaller
than 1.5. } \label{fig:f1}
\end{figure}

We introduce $\alpha(4650)$ (the slope of  the  continuum  around
4650\,\AA) as a new selection criterion.  The reason is that  this
quantity  has been found to be more closely associated with the
recent star formation  history of the galaxy than other quantities
such as  D$_{4000}$  and  high-order  Balmer emission lines. In
Figure~\ref{fig:f1},   we can see that  most of star-forming
galaxies with $\alpha(4650)<-0.001$ have D$_{4000}<1.5$, and all
the W-R galaxies have values of D$_{4000}$ less than 1.5, which
means that the  stellar  populations  of these galaxies have  mean
characteristic ages  less than 1\,Gyr (\cite{Kauffmann03}). Thus
$\alpha(4650)$ is really an indicator of  mean  stellar age, and
hence a suitable criterion for selecting early-stage star bursts.
Also, one would expect that for a cluster with W-R stars, its
color is  dominated  by OB stars, hence blue, so the continuum
around 4650\,\AA\ should be steeper. A detailed description of the
selection procedure will be presented in Zhang et al. (2008, in
preparation), where we have carried out extensive tests to show
that the use of $\alpha(4650)$ does not introduce significant
selection bias into  our W-R sample. For example, we have also
visually examined the spectra of a large number of star-forming
galaxies of  $\alpha(4650)$  $>-0.001$  and found no objects with
obvious W-R features.

We aim to select a sample of W-R galaxies showing evident W-R
features in  their spectra, at least the blue emission bump around
4650\,\AA\ and possibly also a red bump around 5808\,\AA. Our
selection is  based  on  the  data catalogues from the SDSS
studies at  MPA/JHU, publicly available at {\it
http://www.mpa-garching.mpg.de/SDSS/} (see also
\cite{Brinchmann04}). The current version of these catalogues is
based on the SDSS  DR4. We start with all the MPA/JHU objects that
are  classified as {\it star-forming  galaxies}, including both
high  S/N  and  low S/N  ones (see \cite{Brinchmann04} for a
detailed description). The procedure of galaxy classification can
be found in Kauffmann et al. (2003). We then measure the
$\alpha(4650)$ of each galaxy. A total number of 32828 candidates
are then selected from these star-forming galaxies by requiring
$\alpha(4650)$ $\le$ --0.001. Next, we visually examine the
spectra of these candidates and keep those with obvious W-R
features. This resulted in a sample of 866 objects. Since some of
these objects may be separate \hii\ regions in a same galaxy, we
visually examined their $r$-band images and obtained a final
sample of 846 individual W-R galaxies located in the full SDSS DR4
sky region. In some cases, these galaxies show both the blue  and
red bumps, while in other cases only the blue bump is visible. We
noticed that almost all of our W-R galaxies (838/846) are
classified as high S/N star-forming galaxies in the MPA/JHU
catalogues.

We have constructed 20 control samples of non-W-R galaxies from
the underlying star-forming galaxies, by simultaneously matching
five physical parameters: redshift, absolute magnitude,
concentration, D$_{4000}$  and  specific star formation rate. The
matching tolerances are $\Delta  cz  <  500$\,km~s$^{-1}$, $\Delta
M_{^{0.1}r} <  0.3$,  $\Delta  C  <  0.2$, $\Delta$D$_{4000}<0.05$
and $\Delta \log _{10}({\rm SFR}/M_\ast) < 0.5$.

When computing the correlation functions, we need to have complete
knowledge of  the observational selection effects in the NYU-VAGC
release. This is the reason why we select our reference sample
from NYU-VAGC itself. Actually, both the MPA catalogue and the
NYU-VAGC are based on SDSS DR4. The  reason  why we did not match
the two catalogues and find counterparts from  each other is that
our methodology of computing cross-correlation functions does not
require the sample being studied (the W-R or control sample) to be
a subsample of the reference sample.

We used  the  New  York  University  Value  Added  Galaxy
Catalogue (NYU-VAGC)\footnote{\it
http://wassup.physics.nyu.edu/vagc/},
which is also DR4-based and is described in detail in Blanton et
al. (2005), to construct a  reference  sample  of $\sim$300,000
galaxies. The galaxies have $0.01 < z < 0.3$,  $14.5<r<17.6$  and
$-23< M_{^{0.1}r}<-17$, $r$ being the $r$-band Petrosian apparent
magnitude corrected for foreground extinction, and $M_{^{0.1}r}$,
the $r$-band  absolute magnitude corrected to redshift $z=0.1$.
This sample formed  the basis of previous investigations of the
correlation  function, power  spectrum, pairwise velocity
dispersion distributions, and luminosity/stellar mass functions of
galaxies. In order to compute the cross-correlation functions, we
constructed random samples that were meant to include all the
observational selection effects. First, in order to contain the
survey sample, a spatial volume that is sufficiently large is
selected. Secondly, we randomly distribute points within the
volume and eliminate the points that are outside the survey
boundary. Finally,  we select  random galaxies within the same
magnitude limits as the observational sample (\cite{Li06a}). We
also corrected  carefully  for the  effect of   fibre collisions
(\cite{Li06b}).

For quantifying the clustering in the W-R (or matched control) and
reference galaxy samples, we use the two-point correlation
function $\xi(r_p,\pi)$, which measures the excess probability
over random, at separations perpendicular and parallel to the line
of  sight, $r_p$ and $\pi$. $\xi(r_p,\pi)$ is calculated using the
estimator
\begin{equation}
\xi(r_p,\pi)=\frac{N_R}{N_D} \frac{QD(r_p,\pi)}{QR(r_p,\pi)}-1,
\end{equation}
where $N_D$ and $N_R$ are the number of galaxies in the reference
sample and  in the random sample; $QD(r_p,\pi)$ and $QR(r_p,\pi)$
are  the  cross  pair  counts between the W-R (or control) and
reference samples, and between the W-R (or control) and the random
samples, respectively.

Once $\xi(r_p,\pi)$ is estimated, we integrate along the redshift
direction to obtain the projected correlation function,
\begin{equation}
\omega(r_p)=\int ^{+\infty}_{-\infty} \xi(r_p,\pi) d\pi= \sum _i
\xi(r_p,\pi _i) \Delta \pi _i.
\end{equation}
Following Li et al. (2006b), the summation  for  computing
$\omega(r_p)$  runs  from $\pi_1=-39.5 h^{-1}$\,Mpc to
$\pi_{80}=39.5 h^{-1}$\,Mpc,  with  $\Delta\pi_i=1 h^{-1}$\,Mpc.
The errors on the clustering measurements are estimated using  the
bootstrap resampling technique (\cite{bbs84}; \cite{Li06a}).

\section{Results}

\begin{figure}[b]

\vs \centering
\includegraphics[width=0.55\columnwidth]{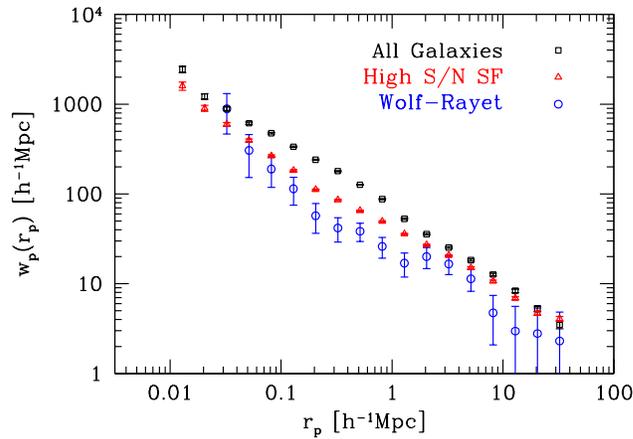}
\caption{\baselineskip 3.6mm  Projected  cross-correlation
function   between   W-R galaxies   and reference galaxies is
plotted as circles.   The cross-correlation  function  of high S/N
star-forming galaxies and  the  same  reference sample is plotted
as triangles. The auto-correlation function of the reference
galaxies is plotted as squares.} \label{fig:f2}
\end{figure}

In Figure~\ref{fig:f2},   we  show  $w_p(r_p)$, the projected
cross-correlation function of the  W-R  galaxies with respect to
the  reference  sample  ({\it circles}).  For comparison, we also
show the cross-correlation function  between the set of the high
S/N star-forming galaxies in SDSS DR4  and the same  reference
galaxies ({\it triangles}), as well as the projected
auto-correlation  function of the reference galaxies ({\it
squares}). Compared  to  the  underlying star-forming galaxies and
the reference galaxies, the W-R  galaxies  are  the least strongly
clustered on scales larger  than  $\sim$0.1 $h^{-1}$Mpc. On
smaller scales, the W-R galaxies seem to cluster more strongly
than the star-forming  galaxies, but the effect is weak.  In
particular, Figure~\ref{fig:f2}  shows that there is a noticeable
change in the slope of the correlation function of W-R galaxies at
$r_p\approx$1 $h^{-1}$Mpc.  Such a change in slope can also be
seen in the curves for the star-forming galaxies and the reference
galaxies, but it  is more pronounced for the W-R galaxies.

\begin{figure}
\centering
\includegraphics[width=0.55\columnwidth]{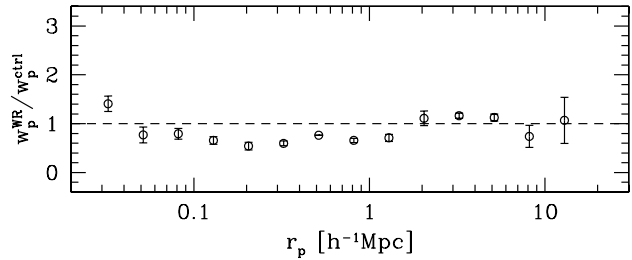}

\caption{\baselineskip 3.6mm
 $w_p(r_p)$ ratio between the W-R
galaxies to the average of 20 control samples of non-W-R
star-forming galaxies, as a function of $r_p$.} \label{fig:f3}
\end{figure}

To eliminate any zeroth-order trends with galaxy mass, structure
and mean stellar age, as discussed in Section~1, we compare the
clustering measurement of the W-R galaxies with that of the
control sample of non-W-R star-forming galaxies. The result is
shown in Figure~\ref{fig:f3},  where  we plot the ratio of the
$w_p(r_p)$ measurement of W-R galaxies to the average measurement
of 20 control samples.  One can see that, on scales larger than  a
few Mpc,  the clustering amplitude of  W-R galaxies does not
differ significantly from  that of  similar, non-W-R galaxies. In
contrast, on the intermediate scales, there is still a difference,
in  the sense  that the ratio between the two cross-correlation
functions exhibits a pronounced `dip' at  scales between
100$h^{-1}$\,kpc and 1$h^{-1}$\,Mpc.  As pointed out by Li et al.
(2006b), the error bars estimated using the bootstrap resampling
technique do not take into account  effects due to cosmic
variance, which can induce significant fluctuations in the
amplitude of the correlation function from one part  of  the sky
to another. We therefore follow Li et al. (2006b) and  divide the
survey into several different areas on the sky, and recompute the
$w_p(r_p)$ ratio for each of these subsamples. We find that, on
scales between 0.1 and 1 $h^{-1}$\,Mpc, all the subsamples, the
ratio (WR to control) lie systematically below unity, indicating
that  the  dip  seen  in Figure~\ref{fig:f3} is robust.  We have
also examined  the dispersion  in the measurement caused by
differences between  the control  samples, and once again there is
a clear indication that W-R galaxies are less strongly clustered
on  these scales relative to all of the control samples.

\begin{figure}
\centering
\includegraphics[width=0.55\columnwidth]{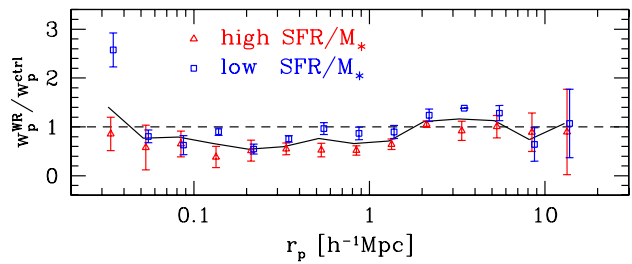}

\caption{\baselineskip 3.6mm Similar  to Fig.\,3  but separately
for W-R galaxies with high and low ${\rm SFR}/M_\ast$. For
comparison, the result for all  W-R  galaxies shown in Fig.\,3 is
plotted here as a solid line.} \label{fig:f4}
\end{figure}

We ordered all the  W-R  galaxies by decreasing ${\rm
SFR}/M_\ast$, then defined the first half as  `high'  SFR objects,
and the second half as `low' SFR objects. In Figure~\ref{fig:f4},
we display the $w_p(r_p)$  ratio curve separately for the high and
low SFR objects. We find there is a difference in  the strength of
the dip at the intermediate scales: the dip is stronger for the
W-R galaxies with higher  star formation rates.

\section{Discussion}

Our  clustering  results  on  large  scales  presented  in
Figure~\ref{fig:f2} demonstrate that W-R galaxies are found in
less massive dark matter haloes in the general population of
star-forming  galaxies.   When  we  compare  the clustering of W-R
galaxies relative to carefully  matched  control samples,  we find
that  the  difference in  the  large scale clustering disappears.
This indicates that W-R galaxies and non-W-R control galaxies
populate  dark  matter haloes of similar masses, which is
consistent with the fact  that  W-R  galaxies are not  a random
subsample  of the  underlying  star-forming galaxies. As mentioned
in Section~1, W-R galaxies are found to be on average less
luminous,  bluer, more concentrated, and younger in mean stellar
age and  higher  in  SFR  than all star-forming galaxies.  This is
why we see  differences  on large scales in Figure~\ref{fig:f2}
where  W-R  galaxies  are compared  to all star-forming galaxies,
but do not see differences in Figure~\ref{fig:f3} where they  are
compared to control samples of similar properties.

Our most interesting result is that, on scales between 0.1  and
1$h^{-1}$\,Mpc, W-R galaxies are significantly
negatively biased
relative to non  W-R  star-forming galaxies of the same luminosity,
concentration, mean stellar  age  and  specific SFR.  We also see
from Figure~\ref{fig:f2} a clear change around  1$h^{-1}\,$Mpc in
the slope of the cross-correlation function  of  W-R  galaxies,  and
such  a change is much more remarkable for W-R galaxies than for all
galaxies  and  all star-forming galaxies.  These two observational
facts  imply  that  W-R  galaxies tend to occupy  preferred positions
within  their  dark  matter  haloes  where conditions are more
favourable for producing massive stars.  We speculate  that such
preferred locations are the centers of
the  dark  haloes, based on the following observational
and theoretical considerations.
\begin{enumerate}

\item The change  in the slope of the correlation  function can
be  understood  as  the transition between the scales where the
pair  counts  are  dominated by  galaxy pairs in the same halo to
those  where  galaxy  pairs are mostly  in  separate haloes
(\cite{Jing04}).

\item The physical scales of 0.1--1\,Mpc, where we do see
significant difference in the clustering of W-R galaxies and
non-W-R galaxies, are comparable  to  the diameters of the dark
matter haloes that are expected  to  host  galaxies  with
luminosities of $\sim L_\ast$ (\cite{Mandelbaum06}).

\item Gas around halo centers could reach high  enough
overdensities to cool via radiative processes. This effect is more
efficient in the less  massive dark matter haloes that are
expected  to  host galaxies  with  stellar  masses comparable to
the objects in our sample.

\item The negative bias observed here for W-R galaxies looks quite
similar  to  that of AGN presented in Li et al. (2006b).   In that
paper,  the authors  used  mock  catalogues constructed from
high-resolution $N$-body  simulations  to  show  that  such  a
negative bias  between  0.1--1$h^{-1}$\,Mpc  can  be   explained
by AGN residing preferentially   at   the    centers    of
their dark matter    haloes.

\item We notice that the behaviour of  the  cross-correlation
function  between W-R galaxies and reference galaxies is very
similar  to  the  cross-correlation function between low-mass
groups and  relatively  bright  galaxies  obtained  by Yang et al.
(2005b).  Since we use a magnitude-limited sample, most of the
reference galaxies are relative bright.

\item There have been studies claiming possible detection of  W-R
stars  in central cluster galaxies (e.g. \cite{Allen95}).
\end{enumerate}

In order to understand this conjecture in more details,  we  have
examined  the distribution  of  W-R  galaxies  more  closely,
using the  SDSS  galaxy  group catalogue of  Yang et al. (2008).
This catalogue was constructed using the halo-based group finder
of Yang et al. (2005a) and  applied to the NYU-VAGC. This group
finder uses the general properties of cold dark matter (CDM)
haloes (i.e.  virial radius, velocity dispersion, etc.)  to
determine  the membership of groups (\cite{Weinmann06}). We search
for counterparts of the 846 W-R galaxies and the control galaxies,
in the group catalogue. Of the W-R galaxies, 257 have been linked
to groups.  We find that $\sim$ 82\% of these  W-R galaxies are
the central galaxies of groups, compared  to $\sim$74\%  for the
corresponding control galaxies. In order to test whether  this
result  suffers from any possible selection effects, we analysed
the clustering properties of both the 257 group members and the
galaxies that are not included  in the group catalogue, using the
same method as described in Section~3. The result showed no
significant differences between these subsamples and  the whole
W-R sample. We should point out that those galaxies that  are  not
in the group catalogue are mostly in low mass haloes, since  the
correlation  function amplitude is slightly lower than that of
those linked to groups. However,  the ratio between W-R galaxies
that are not in groups and the corresponding control galaxies
remains unchanged and hence the dip on intermediate  scales  also
remains unchanged, only because the amplitude of the correlation
function  of  the control galaxies is also reduced.  Therefore,
such a direct look into the observational group catalogue provides
an encouraging  support  to our conjecture described above that
W-R galaxies reside preferentially at  the centers  of  their dark
matter haloes.

\begin{figure}

\vs \centering
\includegraphics[width=0.6\columnwidth]{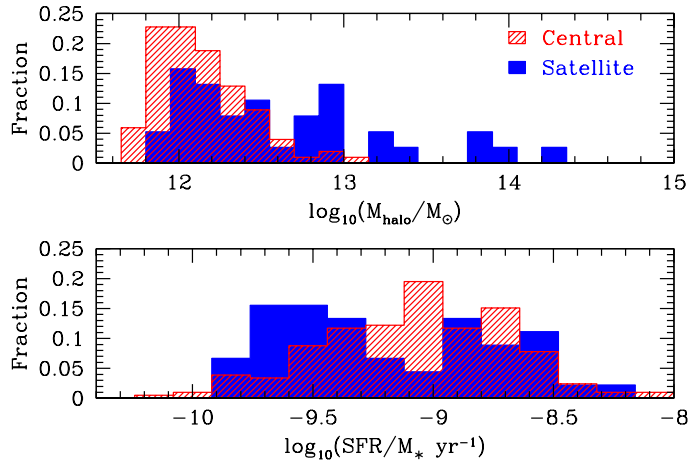}
\caption{\baselineskip 3.6mm  Distributions of the virial mass of
host dark matter haloes ({\it  top})  and
of the specific star formation rate  ({\it bottom}) for  W-R  galaxies  that  are
centrally located ({\it red})  or are satellites  ({\it  blue}) within
their  own groups.} \label{fig:f5}
\end{figure}

We plot in Figure~\ref{fig:f5}  the distribution of the virial
mass of the host dark matter haloes ({\it top}) and the specific
SFR ({\it bottom}) for  W-R  galaxies that are centrally located
({\it red}) or are satellites ({\it blue})  within  their own
groups.  We find that the W-R galaxies  are  hosted,  on average,
by  dark matter haloes of  masses  of $10^{12.3}M_\odot$. The
centrally-located  W-R galaxies, which dominate the W-R galaxy
population, tend to be hosted  by  less massive dark matter haloes
with  a  mean  virial mass  of $10^{12.1}M_\odot$, compared to
$10^{12.7}M_\odot$ for the satellite  W-R  galaxies.   Moreover,
these satellites exhibit a bimodal distribution in SFR, which  is
not  seen  for  the centrally-located W-R galaxies. In addition,
the centrally-located W-R galaxies have higher star formation
rates than the satellite galaxies.  This can  explain  the trend
of clustering with SFR seen in Figure~\ref{fig:f4}, if one accepts
our conjecture that the negative bias of W-R galaxies on
intermediate  scales  is mainly driven by the fraction of
centrally located W-R galaxies.

We would like to point out that the examination based on the group
catalogue is not a strong enough support to our halo-center
conjecture. First, the number of W-R galaxies is really too small
and only a third of them are linked to groups. The fluctuations
in the  fractional  statistics  is  thus  rather  large.   In
addition, the group catalogue itself is incomplete for faint
luminosities that are typical for the  W-R  sample. To  provide
more convincing observational evidence in support of our
conjecture, it needs larger samples and more work  in the future.

We searched for W-R galaxies  only  among  star-forming  galaxies.
There  is  no denying that previous studies have also presented
evidence of W-R  features  in Seyfert 2 and  LINERS (e.g.
\cite{oc82}; \cite{heckman97}). However, what we are interested in
is the ``traditional" W-R galaxies for which the nebular spectrum
is likely due to the photoionization of stellar origin (Schaerer
et al. 1999).
By  this definition, these W-R galaxies are
simply the very early phase of starburst alaxies. Therefore,  by
comparing them and the regular star-forming galaxies in regard to
clustering we will be able to see whether star-forming galaxies at
different locations are at different starburst stages. Our results
here give a positive answer, that the difference in location
between  W-R  galaxies and regular star-forming galaxies may tell
us  something  about merger-triggered starbursts in galaxies.  In
other word, if we think that starbursts are mainly triggered  by
interactions/mergers, then the negative bias of W-R galaxies is
consistent with a picture in which the early phase of starbursts
are more frequently seen in central-satellite mergers, but the
starbursts in satellite-satellite mergers have passed this early
stage and  do  not  show  W-R features any more. The reason why we
do not see in Figure~\ref{fig:f2} the expected rapid rise of
$w_p(r_p)$ on the smallest scales is only that the size of the W-R
sample is too small and that we have no measurements on scales
smaller than $\sim$0.3\,Mpc. From Figure~\ref{fig:f2}, we see that
the correlation of W-R galaxies does exhibit a rapidly rising
tendency on these scales.

\begin{acknowledgements}
We thank Cheng Li for significant input into this project  from
the beginning, and for his help with the methodology of computing
cross-correlation  functions and constructing control samples.  We
are grateful to Frank van den Bosch,  Anna Pasquali, Xiaohu Yang
and Houjun Mo  for  helpful  comments  and  for  providing their
SDSS galaxy  group  catalogue.   We  thank  Yipeng  Jing   and
Andreas Faltenbacher for helpful discussions.  This work is
supported by the National Natural Science Foundation of China,
through Grants 10573014 and 10633020, and the Bairen Project of
the Chinese Academy of Sciences. This work has made use of the
SDSS data and MPA/JHU catalogues of SDSS galaxies. Funding for the
SDSS and SDSS-II has been provided  by  the  Alfred  P. Sloan
Foundation, the Participating Institutions,  the  National Science
Foundation, the U.S.   Department  of  Energy,   the National
Aeronautics and Space Administration, the Japanese Monbukagakusho,
the Max Planck Society, and  the Higher Education  Funding Council
for England.   The  SDSS Web Site is {\it http://www.sdss.org/}.
The SDSS  is  managed  by the Astrophysical  Research Consortium
for the Participating Institutions.  The  Participating
Institutions are the American Museum of Natural  History,
Astrophysical Institute  Potsdam, University of Basel, Cambridge
University, Case  Western Reserve University, University of
Chicago, Drexel University, Fermilab, the Institute  for  Advanced
Study, the Japan Participation Group, Johns Hopkins  University,
the  Joint Institute  for Nuclear Astrophysics,  the   Kavli
Institute for Particle Astrophysics and Cosmology, the Korean
Scientist Group, the Chinese  Academy of Sciences by the LAMOST
Project, Los Alamos National Laboratory,  the Max-Planck-Institute
for Astronomy (MPIA), the Max-Planck-Institute for  Astrophysics
(MPA), New Mexico State University,  Ohio  State University,
University of Pittsburgh, University of  Portsmouth, Princeton
University, the United States Naval Observatory, and the
University of Washington.
\end{acknowledgements}


\begin{thebibliography}{99}
\small \setlength{\itemindent}{-3mm} \setlength{\itemsep}{-0.5mm}
\setlength{\baselineskip}{4.5mm}

\bibitem[Adelman-McCarthy et al. 2006]{Adelman06} Adelman-McCarthy  J.~K.,
Ag{\"u}eros  M.~A., Allam  S.~S. et al., 2006, \apjs, 162, 38

\bibitem[Allen et al. 1976]{Allen76} Allen  D.~A., Wright  A.~E., Goss W.~M.,
1976, \mnras, 177, 91

\bibitem[Allen 1995]{Allen95}   Allen S.~W., 1995,  \mnras, 276,  947

\bibitem[Barrow et al. 1984]{bbs84} Barrow J.~D.,  Bhavsar  S.~P.,
Sonoda D.~H., 1984, \mnras, 210, 19

\bibitem[Blanton et al. 2005]{Blanton05} Blanton  M.~R., Schlegel  D.~J.,
Strauss M.~A. et  al.,  2005,  \aj, 129, 2562

\bibitem[Brinchmann  et  al. 2004]{Brinchmann04}  Brinchmann  J.,  Charlot  S.,
White S.~D.~M. et al., 2004, \mnras, 351, 1151

\bibitem[Conti 1991]{Conti91}   Conti  P.~S.,  1991,  \apj,  377,   115

\bibitem[Guseva et al. 2000]{Guseva00} Guseva  N.~G.,  Izotov  Y.~I.,  Thuan
T.~X., 2000, \apj, 531, 776

\bibitem[Heckman et al. 1997]{heckman97}  Heckman  T.~M.,  Gonzalez-Delgado  R.,
Leigherer C. et al., 1997, \apj, 482, 114

\bibitem[Jing \&  B{\"o}rner 2004]{Jing04}  Jing  Y.~P.,  B{\"o}rner  G.,
2004, \apj, 617, 782

\bibitem[Kaiser 1986]{Kaiser86}  Kaiser  N.,   1986,  \mnras,  219,  785

\bibitem[Kauffmann et  al. 2003]{Kauffmann03}   Kauffmann  G., Heckman T.~M.,
Tremonti  C. et  al.,  2003, \mnras, 346, 1055

\bibitem[Kniazev et  al. 2004]{Kniazev04}  Kniazev   A.~Y.,  Pustilnik   S.~A.,
Grebel  E.~K. et al., 2004, \apjs, 153, 429

\bibitem[Li et al. 2006a]{Li06a} Li  C., Kauffmann  G., Jing    Y.~P.  et  al.,
2006a, \mnras, 368, 21


\bibitem[Li et al. 2006b]{Li06b} Li  C.,  Kauffmann   G.,  Wang   L.   et  al.,
2006b, \mnras, 373, 457


\bibitem[Maeder \& Conti 1994]{Maeder94} Maeder  A.,  Conti   P.~S.,  1994,
\araa, 32, 227

\bibitem[Mandelbaum et  al. 2006]{Mandelbaum06}  Mandelbaum   R.,  Seljak   U.,
Kauffmann  G. et al., 2006, \mnras, 368, 715

\bibitem[Osterbrock \& Cohen 1982]{oc82} Osterbrock  D. E.,  Cohen  R. D.,  1982,
\apj, 261, 64

\bibitem[Peebles 1980]{Peebles80}  Peebles   P.~J.~E.,  1980,
The  Large-Scale Structure  of   the   Universe,  Princeton:
Princeton Univ. Press 


\bibitem[Schaerer  et  al. 1999]{Schaerer99}  Schaerer   D.,  Contini   T.,
Pindao  M., 1999, \aaps, 136, 35

\bibitem[Weinmann et al. 2006]{Weinmann06}  Weinmann    S.~M.,  van  den  Bosch,
Frank C. et al., 2006, \mnras, 366, 2

\bibitem[Yang et al. 2005a]{Yang05a} Yang  X.~H., Mo H.~J.,  van  den  B.  et
al., 2005a, \mnras, 356, 1293

\bibitem[Yang et al. 2005b]{Yang05b} Yang  X.~H., Mo H.~J.,  van  den  B.  et
al., 2005b, \mnras, 362, 711

\bibitem[]{} Yang  X. H., Mo H. J., van den Bosch et al., 2008, ApJ, 671, 153

\bibitem[York et al. 2000]{York00} York  D.~G., Adelman  J., Anderson  J.~E.~J.
et al., 2000, \aj,  120, 1579

\bibitem[Zehavi et al. 2004]{Zehavi04} Zehavi  I., Weinberg  D.~H., Zheng  Z. et
al.,  2004,  \apj, 608, 16

\bibitem[Zehavi et al. 2005]{Zehavi05} Zehavi  I., Zheng  Z., Weinberg  D.~H. et
al.,  2005,  \apj,  630, 1

\bibitem[Zhang et al. 2007]{Zhang07} Zhang  W., Kong  X., Li  C., Zhou   H.~Y.,
 Cheng  F.~Z., 2007, \apj, 655, 851
\end{thebibliography}
\end{document}